\documentclass[twocolumn,footinbib,superscriptaddress,floatfix,bibliography]{revtex4-2}
\usepackage{amsmath}
\usepackage{amssymb}
\usepackage{physics}
\usepackage{graphicx}
\usepackage{hyperref} 
\usepackage{pgfplots}
\usepackage[utf8]{inputenc}
\usepackage[T1]{fontenc}
\usepackage{textcomp}
\usepackage{ulem}
\usepackage[caption=false]{subfig}		

\newcommand{\nic}{\textcolor{black}}

\begin{document}
	\title{Witnessing Disorder in Quantum Magnets}
	
	\author{Snigdh Sabharwal}
	\email{snigdh.sabharwal@oist.jp}
	\affiliation{Theory of Quantum Matter Unit, Okinawa Institute of Science and Technology Graduate University, Onna-son, Okinawa 904-0412, Japan}
	
	\author{Tokuro Shimokawa} 
	\affiliation{Theory of Quantum Matter Unit, Okinawa Institute of Science and Technology Graduate University, Onna-son, Okinawa 904-0412, Japan}
	
	\author{Nic Shannon}
	\affiliation{Theory of Quantum Matter Unit, Okinawa Institute of Science and Technology Graduate University, Onna-son, Okinawa 904-0412, Japan}
	
	\date{\today}
	\begin{abstract}
		There are no clean samples in nature.  Therefore, when we come to discuss the entanglement properties of quantum materials, the effects of disorder must be taken into account.  This question is of particular interest for high-entangled phases, such as quantum spin liquids, which lie outside the Landau paradigm for classifying phases of matter. In this work, we explore what experimentally-accessible measures, in the form of concurrence, residual tangle  and quantum Fisher information, can teach us about the entanglement in the presence of disorder.  As a representative example, we consider the Tomonaga-Luttinger liquids (TLL) and disorder-driven random singlet (RS) phases found in antiferromagnetic quantum spin chains.  Using quantum Fisher information  and residual tangle, we demonstrate that both TLL and RS phases exhibit multi-partite entanglement.  In the case of the RS phase, we attribute this to entanglement localized below a crossover length scale. We further show that the order of disorder average matters in calculating measures like concurrence, and that this can lead to false inferences when interpreting experiment.  Nonetheless, correctly interpreted, these witnesses provide useful information about the effects of disorder.  We explore how information about the central charge of the TLL can be extracted from the low-temperature behavior of concurrence, and conjecture that this analysis can be extended to the effective central charge of the RS phase.  Finally, we establish how RS and TLL phases can be distinguished through the growth of multi-partite entanglement, as witnessed by the equal-time structure factor. These results establish that, used carefully, experiments based on entanglement witnesses can provide important information about quantum spin systems in the presence of disorder.
	\end{abstract}
	\maketitle
	\section{Introduction}
	How does disorder affect entanglement within quantum materials? Can we quantify such effects experimentally?
	In nature, all materials possess a certain degree of disorder.  Thus, when trying to understand any phase of matter, the role of disorder is an important factor to take into consideration. Disorder or randomness is especially problematic when trying to  identify quantum spin liquid (QSL) phases within candidate materials. This question arises in the context of many different materials, including  $\kappa$-ET$_{2}$Cu$_2$(CN)$_3$~\cite{Shimizu2003,Kurosaki2005,yamashita2008thermodynamic,yamashita2009thermal,pratt2011magnetic,Watanabe2014,Shimokawa2015,Wu2019}, Ba$_6$Y$_{2}$Rh$_2$Ti$_2$O$_{17-\delta}$ \cite{Nguyen2021,Wonjun2024}
	, ZnCu$_3$(OD)$_6$Cl$_2$ (Herbertsmithite)~\cite{Shores2005,Helton2007,han2012fractionalized,Rajiv2010,Kawamura2014,fu2015evidence,Shimokawa2015,Kawamura2019,khuntia2020gapless,wang2021emergence,Quentin2022}, YbMgGaO$_4$~\cite{Li2015,Zhu2017,paddison2017continuous,Kimchi2018}, H$_3$LiIr$_2$O$_6$~\cite{kitagawa2018spin, Knolle2019,Lee2023}, etc. Here  the challenge lies in being able to distinguish quantum fluctuations from disorder-induced ones. 
	This puts into question whether it is the QSL or some disorder induced phase realized within these materials?\\
	
	Entanglement is a valuable resource to identify phases such as QSLs \cite{Broholm2020}. Intuitively, one expects disorder, which localizes correlations, to have an effect on entanglement. The key issue, however, is whether this information can be  accessed experimentally.  Unlike cold atom systems where non local measures like (R\'{e}nyi) entanglement entropy are accessible~\cite{Islam2015,Kaufman2016}, the same does not extend to condensed matter systems.
	Consequently, only limited information about the underlying phase can typically be accessed experimentally. It is thus important to determine if useful information about the entanglement structure can be obtained within experimental constraints.  Can we do so by making use of experimentally-accessible entanglement measure~\cite{Wiesniak_2005,Brukner2006,Rappoport2007,Souza2008,Souza2009,soares2009entanglement,Horodecki2009a,GUHNE20091,Candini2010,Yurishchev2011,reis2012evidence,Hyllus2012,Das_2013,Singh_2013,aldoshin2014quantum,chakraborty2014signature,chakraborty2015experimental,singh2015experimental,sahling2015experimental,Hauke2016,garlatti2017portraying,garlatti2019unravelling,Kon2019,Mathew2020,Scheie2021,Pratt2022,Laurell2021,athira2023bipartite,Scheie2023,scheie2023reconstructingspatialstructurequantum,Laurell2025}, for instance?
	\\~\\
	In this study, we explore how measures based on entanglement witnesses can be used to distinguish the different phases which arise in quantum spin systems with and without disorder. As a representative example, we consider the $S = 1/2$ Heisenberg (XXX) spin chain with and without bond disorder. The competing phases in this case are the Tomonaga Luttinger liquid (TLL) \cite{FDMHaldane_1981} found in the absence of disorder, 
	and the random singlet (RS) phase realised for arbitrarily weak bond disorder~\cite{Ma1979,Dasgupta1980,Fisher1994, Laflorencie2016}.  
	We consider three different measures, namely concurrence  \cite{Hill1997,Wootters1998}, tangle \cite{Coffman2000} and quantum Fisher information (QFI) \cite{Hyllus2012,Hauke2016}, all of which have previously been discussed in the context of experiment \cite{Scheie2021,Laurell2021, Scheie2023, Laurell2025}, and 
	use exact diagonalization to explore how each of these behaves in the TLL and RS phases. 
	\\~\\
	From our analysis, we conclude:
	\begin{itemize}
		
		\item Both the TLL and RS phases are multi-partite entangled. In the TLL, multi-partite entanglement is present at all length scales. However, in the RS phase, multi-partite entanglement is restricted to clusters smaller than a crossover length-scale, beyond which the effects of disorder become relevant. 
		
		\item  Convex roof functions, like concurrence, are sensitive to the order in which disorder-averages are calculated.  In practice, this means that the distribution of concurrence in real space cannot be used to distinguish TLL from RS in experiments on systems with quenched disorder.   		 
		
		\item 
		Information about the central charge of the TLL can be extracted from the low-temperature behaviour of concurrence. We conjecture that this analysis can be extended to the effective central charge of the RS phase, and show how this approach could be used to distinguish TLL from RS in the presence of quenched disorder.
		
		\item The temperature--dependence of multi-partite entanglement, 
		as witnessed by quantum Fisher information (QFI), provides a robust
		means of probing the effect of disorder on quantum spin systems.   
		In the case of quantum spin chains, this provides a clear distinction between the TLL and RS phases which can be accessed in experiment.

	\end{itemize}
	
	For our investigations, we employed exact diagonalization (ED) and the thermal pure quantum (TPQ) methods \cite{imada1986quantum,Hams2000,Sugiura2012,Sugiura2013}. Moreover, throughout this Article we work in the units where $\hbar = 1$ and $ k_{B} = 1$. \\
	
	 For completeness, we note that concurrence at zero temperature has previously been discussed in the context of disordered spin chains \cite{Hoyos2006,Mohdeb2020}. An interesting alternative approach distinguishing RS from the TLL phase employing Bell's inequality was discussed in \cite{Hoyos2006}.\\
	
	The remainder of the Article is structured as follows. In Sec.~\ref{sec:EntanglBackgrnd}, we start out by giving a brief background on how one can access and/or detect entanglement  in a limited setting using only one and two point spin-correlation functions. In our work, we employ concurrence to characterize the pairwise entanglement, residual tangle (RT) to detect the presence of multi-partite entanglement and the QFI to characterize the depth of multi-partite entanglement within the state. The subtleties, where they arise, are explained along the way. 
	
	In Sec.~\ref{sec:Notation}, we detail our notation for evaluating averages. Sec.~\ref{sec:HAFChainOverview} reviews key results related to the two states under investigation: the TLL and RS phases.  Sec.~\ref{sec:Problem} introduces the model, disorder, and parameters used to investigate the entanglement structure of these phases. Following which, in Sec.~\ref{sec:ZeroTemp}  we present our zero temperature results for concurrence, RT and QFI.  Sec.~\ref{sec:FiniteTemp} explores the finite temperature behavior of concurrence and equal time structure factor as witnesses. Here we address the subtleties involved with disorder averaging when dealing with convex roof measures like concurrence. We also discuss how to access the central charge for these phases using concurrence and demonstrate the utility of using a multi-partite entanglement witness like equal-time structure factor as a probe for observing disorder effects.
	
	In Sec.~\ref{sec:Disc}, we discuss at length the nature of entanglement in the RS phase, explaining the observed multi-partite entanglement and its origins. Finally, we conclude our findings in Sec.~\ref{sec:Conclusion}.
	
	\section{Accessing Entanglement in many body quantum systems}\label{sec:EntanglBackgrnd}
	Entanglement in the modern perspective is thought of as a resource, one that can be used for quantum computation and information processing \cite{Lo1998}.  To this end, an important goal within quantum information is to quantify how entangled a state is?  Unfortunately doing so is quite challenging and aside from a few special cases there is no solution as of yet. Even figuring out, given a density matrix, whether a state is separable or entangled, computationally, turns out to be very complex: in fact it is an \textit{NP-hard} problem  \cite{Gurvits2003}. 
	
	On the other hand, when it comes to quantifying or detecting entanglement in experimental settings, especially in condensed matter, the challenge diverges from computational complexity to practical feasibility. With a large number of particles involved, a complete state reconstruction by techniques like quantum state tomography is not possible. Thus quantifying entanglement using entanglement entropy Eq.~(\ref{VNent}),
	\begin{equation}\label{VNent}
		E(\ket{\phi_{AB}}) = S(\rho_{A})  = -\Tr \rho_{A} \log(\rho_{A}) \; , 
	\end{equation}
	where $\rho_{A}$ corresponds to a reduced density matrix obtained by tracing out subsystem $B$ 
	\begin{equation}
		\rho_{A} = \Tr_{B}(\ketbra{\phi_{AB}}) \; ,
	\end{equation}
	is not an option.
	
	So how does one quantify or detect entanglement in a way that is experimentally tractable for condensed matter systems? One strategy is to employ \textit{entanglement witnesses} \cite{Horodecki2009a,GUHNE20091}, which are observables that provide a means to detect entanglement in situations where one possesses limited information about the state. These observables violate certain bounds to indicate the existence of a bi-partite or multi-partite entanglement. 
	
	In what we follows, we briefly review the concepts and definitions needed to understand this Article.
	We start by introducing the concept of entanglement, and then introduce three entanglement witness-based measures: concurrence \cite{Hill1997, Wootters1998}, tangle \cite{Coffman2000}, and quantum Fisher information (QFI) \cite{Hyllus2012, Hauke2016}.
	The advantage of using these measures is that they can be measured from spin correlations and structure factors which are readily accessible by techniques like inelastic Neutron scattering (INS).
	\subsection{Entanglement essentials}\label{EntEss}
	A bi-partite mixed state $\rho$ is said to separable if it can be expressed as 
	\begin{equation}
		\rho = \sum_{i}p_{i}\rho^{A}_{i}\otimes \rho^{B}_{i}\;,
	\end{equation}
	and entangled otherwise. When $A$ and $B$ comprise of a single element, we use the term 2-partite for clarity. For example, $A$ and $B$ could correspond to $S =1/2$ particles on a lattice site. The product state 
	\begin{equation} 
		\rho = \rho^{A} \otimes \rho^{B} \;,
	\end{equation}
	is a special case of the separable state.
	
	The classification of multi-partite states is slightly more subtle since there are many inequivalent ways to partition the system \cite{Hyllus2012,Pezze2014}. 
	Let us begin by imagining a pure state comprised of $N$ particles which we partition into $M$ blocks. Now, if all the blocks are of size at most $k$, then the state is termed \textit{$k$-producible} i.e.,
	\begin{equation}
		\ket{\psi_{\text{k-prod}}} = \otimes_{l=1}^{M} \ket{\psi_{l}} \; \text{such that} \; N_{l} \leq k \;\forall l\;,
	\end{equation} 
	where it is understood
	\begin{equation}
		\sum_{l} N _{l} = N\;.
	\end{equation}
	A state is \textit{$k$-partite entangled} if it is $k$-producible but not $(k-1)$-producible which translates to 
	\begin{equation}
		\ket{\psi_{\text{k-ent}}} = \otimes_{l = 1}^{M} \ket{\psi_{l}} \; \text{such that} \; N_{l}=k \; \text{for some } l\;,
	\end{equation}  
	i.e., there exists a non-factorizable state $\ket{\psi_{l}}$ with exactly  $N_{l}=k$ particles. A mixed state $\rho_\text{k-prod}$ is said to be $k$-producible if it can be written as a convex mixture of $k$-producible pure states, 
	\begin{equation}\label{eq:kprodmix}
		\rho_\text{k-prod} = \sum_{r}p_{r}\ketbra{\psi^{r}_\text{k-prod}} \;\; (k_{l} \leq k \;\; \forall k)\;,
	\end{equation} 
	where
	\begin{equation}
		p_r > 0 \text{ and } \sum_{r}p_r =1\;.
	\end{equation}
	
	Having introduced the necessary terminology, we are in a good position to introduce the entanglement measures. We focus our investigation towards those suited for $S = 1/2$ particles. There do exist generalizations of the measures discussed for arbitrary spin quanta but we do not discuss them here. An exception is the QFI that we discuss in Sec.~\ref{QFI}
	
	\subsection{Concurrence}\label{ConcSec}
	Concurrence quantifies the pairwise entanglement between two $S = 1/2$ particles, 
	\nic{by quantifying the ``entanglement of formation'' \cite{Bennett1996}, required 
	to construct a given state}.
	This could be in a system of only two or many $S=1/2$ particles. For a mixed state $\rho_{ij}$ shared between them where $i$ and $j$ are the indices pointing to the particles in question, it was shown \cite{Hill1997, Wootters1998} to have the following relationship
	\begin{equation}\label{conc}
		C_{ij} \equiv C(\rho_{ij}) = \max\{0,\lambda_{1}-\lambda_{2}-\lambda_{3}-\lambda_{4}\} \;,
	\end{equation}
	Here, $ \lambda$'s correspond to the square root of the eigenvalues of a non-Hermitian matrix 
	\begin{equation}\label{def:lambda}
		\rho_{ij} \tilde{\rho}_{ij} \ket{v_{\lambda}} = \lambda^{2} \ket{v_{\lambda}}\;, 
	\end{equation}
	arranged in descending order, with  
	\begin{equation}\label{def:matrix}
		\tilde{\rho}_{ij}  = (\sigma^y \otimes \sigma^y) \rho^{*}_{ij}  (\sigma^y \otimes \sigma^y)\;,
	\end{equation}
	being a spin-flipped state and $ *$ denoting the complex conjugation operation performed in the computational basis. 
	
	Following this definition, concurrence ranges from 0 corresponding to a separable state to 1, a maximally entangled Bell state. It quantifies the degree of 2-partite entanglement regardless of whether the state is pure or mixed. Thus, it remains a good measure of entanglement even at finite temperature. The practical advantage of using concurrence is that it is related to spin correlations \cite{Ulrich2003,Amico2004} and thus a very useful tool to access the 2-partite entanglement in the system. 
	For instance, a SU$(2)$ symmetric system, a $S = 1/2$ spin chain Eq.~(\ref{HamSpinChain}) for example, is given by \cite{Amico2004}
	\begin{equation}\label{concfromspin}
		C_{ij} = 2\max\bigg\{0,  2 |g^{zz}_{ij}| - \bigg|\frac{1}{4}+g^{zz}_{ij}\bigg|\bigg\}\;,
	\end{equation}
	where
	\begin{equation}
		g^{zz}_{ij} = \langle S^{z}_{i}S^{z}_{j}\rangle \;.
	\end{equation}
	In a system with more than two particles one should note that vanishing concurrence doesn't imply absence of higher partite entanglement. Often, one also uses the \textit{two-tangle} \cite{Coffman2000,Roscilde2004,Osborne2006} 
	\begin{equation}
		\tau_{2,i} = \sum_{j\neq i} C^{2}_{ij}\;,
	\end{equation} as a measure of pairwise entanglement which characterizes all possible pairwise entanglement within a many partite system.
	
	\nic{
	As a parting comment, we note that in the case where a pair of spins exist in a 
	pure state, 
	the definition of concurrence reduces to a simpler form 
	\cite{Hill1997}.
	However when considering states with multipartite entanglement, it is necessary 
	to use the more general definition, Eq.~(\ref{conc}), since the reduced density matrix 
	for a pair of spins will not, in general, be a pure state.
	%
	%
	This point is addressed in greater detail Appendix~\ref{appendix:worked.example}.
	}
	
	\subsection{Residual Tangle}\label{residtang}
	Residual tangle (RT) \cite{Coffman2000,Osborne2006,Laurell2025} is a measure that accesses entanglement beyond the pairwise setting, in other words, a multi-partite measure. It is defined as the difference between one-tangle and the sum of pairwise entanglement measured via the two-tangle,
	\begin{equation}\label{eq:RT}
		\tau_{3,i} = \tau_{1,i} - \tau_{2,i}\;,
	\end{equation}
	where 
	\begin{equation}
		\tau_{1, i} = 1-4 \sum_{\alpha = \{x,y,z\}} \langle S^{\alpha}_{i} \rangle^{2}\;,
	\end{equation} 
	is the one-tangle \cite{Coffman2000,Osborne2006,Laurell2025}. 
	
	%
	Strictly, $\tau_{3,i}$ is only well defined for pure states \cite{Laurell2025}.
	Consequently, the interpretation of RT as an entanglement measure becomes inapplicable at finite temperatures. 
	Moreover, although a nonzero RT suggests the presence of multi-partite entanglement, a zero RT does not necessarily indicate its absence.
	
	\subsection{Quantum Fisher Information}\label{QFI}
	The measures discussed previously are very useful, however they come with the 
	limitations of being unable to access multi-partite entanglement, or become invalid as measures at finite temperatures. 
	A measure that does take care of both these issues is the quantum Fisher information (QFI). 
	For a mixed state described using the spectral decomposition 
	\begin{equation}\label{eq:specdecomp}
		\rho = \sum_{k}p_{k}\ketbra{k}\;,
	\end{equation}
	such that $p_{k} >0$ and $\sum_{k} p_{k} = 1$, the QFI is specified, with respect to some operator $\hat{O}$, as
	\begin{equation}
		F_{Q}[\rho,\hat{O}] = 2 \sum_{k,k'}\frac{(p_{k} - p_{k'})^{2}}{p_{k}+p_{k'}} |\langle  k|\hat{O}|k' \rangle|^{2}\;,
	\end{equation}
	which for pure states simplifies to 
	\begin{equation}
		F_{Q}[\ket{\psi},\hat{O}] =  4 (\Delta \hat{O})^2 = 4  \bigg(\langle \hat{O}^2 \rangle - \langle \hat{O}\rangle^2\bigg) \;.
	\end{equation}
	Thus, for pure states, it simplifies to the variance of $\hat{O}$ up to proportionality. At finite temperature, for a system in thermodynamic equilibrium, the probability $p_k$ Eq.~(\ref{eq:specdecomp}) corresponds to the occupation probability as ascribed by the canonical ensemble
	\begin{equation}
		p_{k} = \exp(-E_{k}\beta)/Z \;,
	\end{equation}
	where $Z$ corresponds to the partition function and $\beta$ is the inverse temperature. For this case, the QFI density 
	\begin{equation}
		f_{Q} = F_{Q}/N \;,
	\end{equation}
	can be accessed
	\begin{equation}
		f_{Q}( \beta) =  \frac{4}{\pi}\int_{0}^{\infty} d\omega\tanh(\frac{\omega\beta}{2})\chi''(\beta,\omega)\;,
	\end{equation}
	from $\chi''(\beta,\omega)$, the imaginary, dissipative part of dynamic susceptibility \cite{Hauke2016}, $\chi(\beta,\omega)$
	defined as,
	\begin{equation}
		\chi(\beta,\omega) = \frac{i}{N}\int_{0}^{\infty} dt e^{i\omega t} \tr\bigg(\rho[\hat{O}(t),\hat{O}] \bigg)\;,
	\end{equation}
	with $\beta$ being the inverse temperature.
	
	The connection between QFI and multi-partite entanglement holds \cite{Hyllus2012} for local operators $\hat{O}$ with bounded spectrum. For instance, consider a scenario where $\hat{O}$ is composed of a sum of one-body operators $\hat{o}$ 
	\begin{equation}
		\hat{O} = \sum_{j = 1}^{N} \hat{o}_{j}\;,
	\end{equation}
	with an associated spectral width
	\begin{equation}
		\Delta_{o} = o_{\text{max}} - o_{\text{min}}\;,
	\end{equation}
	where $ o_{\text{min}} $ and $ o_{\text{max}}$ correspond to minimum and maximum eigenvalues of $\hat{o}$. Then the operator, $\hat{O}$ is said to have a bounded spectrum if the spectral width $\Delta_{o}$ is finite.
	For such operators, the QFI is bounded from above for $k$-producible states,
	\begin{equation}
		F_{Q}[\rho_\text{k-prod}, \hat{O}] \leq \Bigg\{ \bigg\lfloor \frac{N}{k} \bigg\rfloor k^2 + \bigg(N - \bigg\lfloor \frac{N}{k}  \bigg\rfloor k\bigg)^2 \Bigg\} \Delta_{o}^2\;.
	\end{equation}
	In the special case when $k$ is a divisor of $N$, this condition simplifies to 
	\begin{equation}
		f_{Q} \leq k\Delta_{o}^2\;,
	\end{equation}
	An assumption is made in situations when $N$ is very large i.e. in the thermodynamic limit then every $k$ is treated as a divisor of $N$. 
	Thus, in the event the QFI density violates the bound above, we can say that underlying state has at least $(k+1)$-partite entanglement i.e.,  has a minimum entanglement depth of ($k+1$),
	\begin{equation}
		f_{Q} > k \Delta_{o}^2 \implies \rho\; \text{is at-least} \;(k+1)\text{-partite entangled}\;.
	\end{equation}
	
	In preparation for what's coming, we will work the following operator
	\begin{equation}
		\hat{O}(q) = \sum_{r} e^{iqr} \hat{S}^{z}(r)\;,
	\end{equation}
	where $r$ is a discrete lattice site index and $q$ is the discrete momentum. By using the fluctuation-dissipation theorem (FDT), the dissipative part of the dynamical susceptibility $\chi''$ can be related to the dynamic structure factor $S^{zz}(q,\omega)$~\cite{lovesey1984theory},
	\begin{equation}
		\chi''(q,\beta,\omega) = \pi (1 - e^{-\beta\omega}) S^{zz}(q,\omega,\beta)\;,
	\end{equation}
	a quantity accessible in INS.
	
	For temperature below some threshold value $T < T_{Q}$, it was shown \cite{Menon2023} that the equal time structure factor, using it's connection to the QFI, can also be used as multi-partite entanglement witness,
	\begin{equation}
		4S^{zz}(q,\beta)  > k \implies \rho\; \text{is at-least}\; k\text{-partite entangled}\;,
	\end{equation}
	where $S^{zz}(q,\beta) $ is the equal time structure factor
	\begin{equation}
		S^{zz}(q,\beta)  = \int_{-\infty}^{\infty}d\omega S^{zz}(q,\omega,\beta)\;.
	\end{equation}
	Note that at zero temperature, for a pure state, the equal time structure factor is proportional to the QFI density
	\begin{equation}\label{eq:fqfromequalstructfact}
		f_{Q}(q) = 4 S^{zz}(q) \;.
	\end{equation}
	
	\section{Notation}\label{sec:Notation}
	With many averages and sums being taken, we would like to bear out the notation here for ease of reading the plots. 
	To evaluate the disorder-average over M random realizations within an ensemble we consider
	\begin{equation}\label{disavg}
		\bar{\Gamma} = \frac{1}{M} \sum_{m = 0}^{M-1}\Gamma^{(m)}\;,
	\end{equation}
	where $\Gamma$ can correspond to some observable
	\begin{equation}
		\Gamma^{(m)}\equiv \langle \Gamma \rangle_{m}
	\end{equation}
	or an entanglement measure discussed in Sec.~\ref{sec:EntanglBackgrnd}. 
	For measures with explicit site dependence, we also perform an average over sites
	\begin{equation}\label{disordavg}
		\bar{\Gamma}_{n} = \frac{1}{MN} \sum_{i = 0}^{N-1} \sum_{m = 0}^{M-1}\Gamma^{(m)}_{ii+n}\;,
	\end{equation}
	represents the disorder-averaged observable for the $n$-th neighbor of the $N$-site spin chain. For example $\Gamma_{ij}$  can be thought to depict the two-point spin correlation or the pairwise concurrence.
	
	\begin{figure*}[t]
		\centering
		\subfloat[]{\includegraphics[width=1\columnwidth]{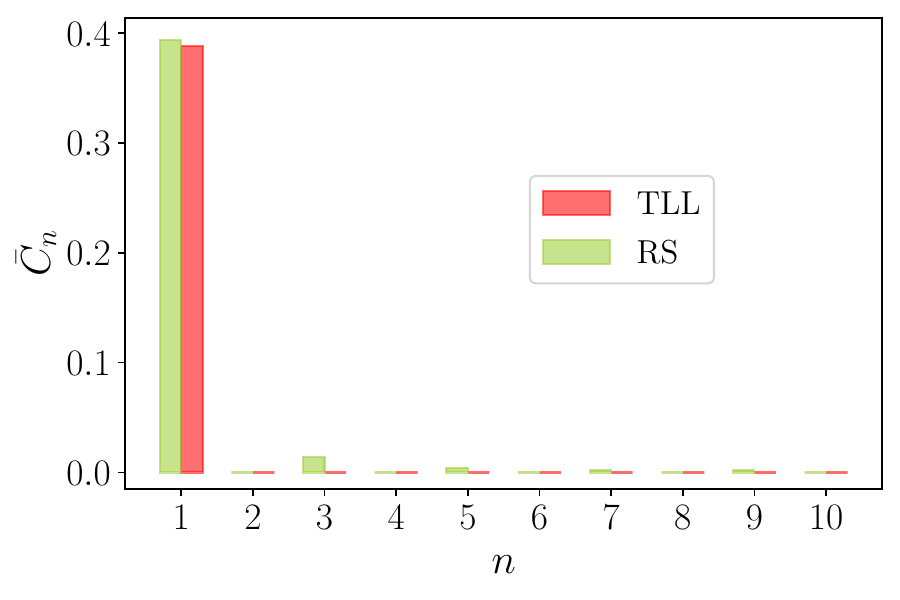} \label{fig:ConcT=0_a}}
		\quad
		\subfloat[]{\includegraphics[width=1\columnwidth,height=0.66\columnwidth]{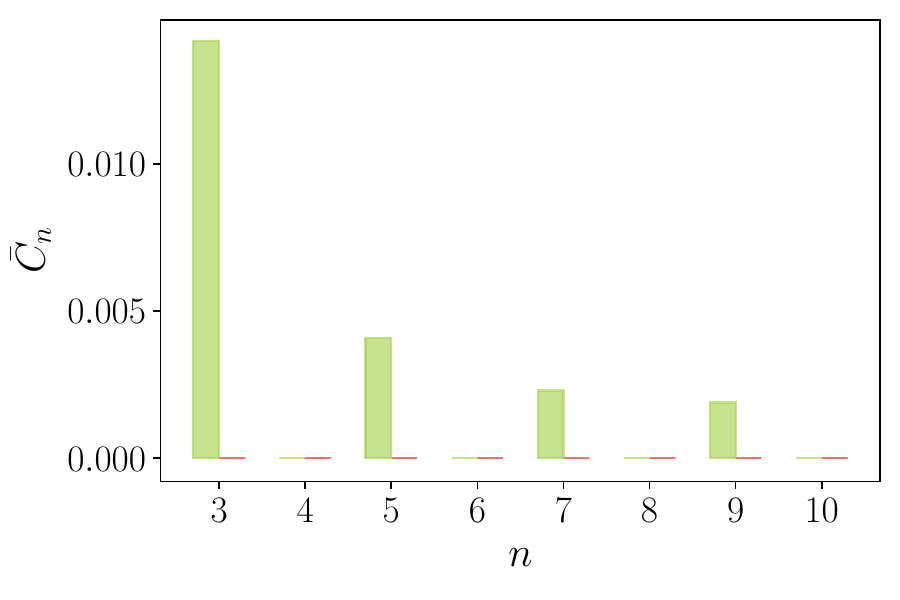} \label{fig:ConcT=0_b}}
		\caption{
		\nic{
		Comparison of distribution of concurrence as a function of separation of spins 
		within the Tomonaga Luttinger liquid (TLL) and random singlet (RS) phases. 
		(a) Distribution of concurrence $C_{ij}$ on $n^{th}$ neighbour bonds, for $n =1\ldots 10$.
		Within the TLL (orange bars), concurrence vanishes identically for all bonds with $n > 1$.
		In contrast, within the RS phase (green bars), concurrence takes on a finite value for all odd values of $n$.
		(b) Detail of $C_{ij}$ for $n>2$, showing slow decay of concurrence within the RS phase.
		All results are shown are taken from exact diagonalization of an $N=30$ spin chain, 
		with $C_{ij}$ defined through Eq.~(\ref{concfromspin}), and disorder averages 
		calculated for $M=1001$ instances, using Eq.~(\ref{disordavg}), with $\Gamma = C_{ij}$.
		For ease of viewing,  error bars have been suppressed.}
		}	
		\label{ConcT=0}
	\end{figure*} 
	
	\section{Overview of Heisenberg Spin chains}\label{sec:HAFChainOverview}
	In this section we will briefly summarize necessary background pertaining to the 
	Tomonaga Luttinger Liquid (TLL) and random singlet (RS) phase in $d = 1$.  
	In this article, by TLL we refer to the long length scale theory of the Heisenberg (i.e. XXX) spin chain 
	characterized by the Luttinger parameter, $K = 1/2$ \cite{Giamarchi2003}. 
	On the other hand, we interpret the strong disorder renormalization group (SDRG) \cite{Fisher1994} 
	as the long length-scale theory of the RS phase. 
	
	We consider the Heisenberg Hamiltonian with anti-ferromagnetic (AF) interactions
	\begin{equation}\label{HamSpinChain}
		H = \sum_{i=0}^{N-1}J_{i}\ \mathbf{S}_{i}\cdot \mathbf{S}_{i+1}\;,
	\end{equation}
	where $J_{i}$ corresponds to the coupling strength between nearest neighbors, 
	and $\mathbf{S}_{i}$ are the usual $S = 1/2$ vector operator at site $i$. 
	
	The isotropic case corresponds to the scenario when all the couplings are equal i.e.,
	\begin{equation}
		J_{i} = J\; (>0) \;\; \forall i\;.
	\end{equation}
	In this case, the model realizes a $SU(2)$ symmetric ground state, whose long wavelength behavior 
	provides an example of a TLL phase, with correlation function  \cite{Affleck1998,LUKYANOV1998533}
	\begin{equation}\label{eq:TailTLL}
		\lim_{|i-j|\to \infty} g^{zz}_{ij} \sim \frac{(-1)^{|i-j|} \sqrt{\log(|i-j|)}}{2\pi^{3/2} |i-j|^{\eta_{\sf TLL}}} \;, 
	\end{equation}
	showcasing a power-law decay with exponent 
	\begin{equation}
		\eta_{\sf TLL} = 1\;.
	\end{equation}
	specific to $K=1/2$. 
	We note that the logarithmic corrections in Eq.~(\ref{eq:TailTLL})  arise due to the irrelevant operators 
	that become marginal for the isotropic Hamiltonian \cite{Affleck1998,LUKYANOV1998533}
	
	We now turn to the case where the couplings $J_{i}$ in Eq.~(\ref{HamSpinChain}) are positive, and randomly drawn from some 
	probability distribution $P(J)$.
	In this case one realizes a RS phase whose long-wavelength is well characterized by the SDRG framework \cite{Fisher1994}.
	Disorder-averaged correlation functions decay algebraically 
	\begin{equation}\label{eq:TailRS}
		\lim_{|i-j|\to \infty} \bar{g}^{zz}_{ij} \sim  \frac{(-1)^{|i-j|} }{ |i-j|^{\eta_{\sf RS}}}\;,
	\end{equation}
	with exponent
	\begin{equation}\label{eq:DecayExpRS}
		\eta_{\sf RS} = 2\;,
	\end{equation}
	independent of the choice of $P(J)$. 
	We note that logarithmic corrections to power-law decay, Eq.~(\ref{eq:TailRS}),  have been reported \cite{YuRong2016}. 
	However, their validity has been questioned in subsequent studies \cite{Getelina2020correlation,Wada2022}.
	
	\section{Setting up the problem}\label{sec:Problem}
	In this Section we make concrete our problem. Our goal here is to analyze what the previously introduced measures inform about the nature of entanglement of the TLL and RS phases, respectively and whether one can use them in a meaningful way to distinguish between the two phases. 
	We investigate the antiferromagnetically interacting Heisenberg Hamiltonian in Eq.~(\ref{HamSpinChain}), 
	with interactions drawn from a box distribution
	\begin{equation}\label{eq:BoxDist}
		P(J) = \begin{cases}
			\frac{1}{2\Delta} , &\;\;\text{if} \;J\in(1-\Delta, 1+\Delta]\\
			0, &\text{otherwise}\;,
		\end{cases}
	\end{equation} 
	where $\Delta$ parametrizes the strength of disorder. 
	For our investigations, we set 
	\begin{eqnarray}
		\Delta &=& 1  \; [\text{disordered}] \; , \nonumber\\
		\Delta &=& 0 \; [\text{clean}] \; .
	\end{eqnarray}	
	
	We make use of exact diagonalization for our analysis and use the quantum typicality method for finite temperature investigations \cite{Sugiura2012,Sugiura2013}. Throughout our analysis we restrict ourselves to spin chains of even length with periodic boundary conditions.
	For $T = 0$ investigations we focused on $M = 1001$ disorder realizatons for $N$ ranging from $16$ to $30$. For $T \neq 0$, we consider $M = 200$ disorder realizations for $N = 16$ and $24$. For $N = 20$, we consider $M = 100$ disorder realizations. In addition, we have taken into account $100$ thermal pure quantum (TPQ) state samples for all three system sizes. 
	
	\section{Zero Temperature results}\label{sec:ZeroTemp}
	In what follows we will discuss the entanglement structure of TLL and RS phases as characterized by concurrence, RT and QFI at zero temperature. 
	
	\subsection{Nearest neighbor and beyond}\label{ConcT=0RSvTLL}
	We start by looking at the distribution of pairwise entanglement in the two  phases. In Fig.~\ref{ConcT=0}, we explore how the concurrence is distributed on the spin chain. For the RS phase, we calculate the disorder-averaged concurrence i.e., we evaluate the $g^{zz}_{ij}$ correlation function for each realization and then from it obtain the corresponding concurrence using Eq.~(\ref{concfromspin}). Thereafter we average over the concurrence per disorder realization as per Eq.~(\ref{disordavg}). For the TLL, we evaluate the sum in Eq.~(\ref{disordavg}) without the disorder-average. 
	
\begin{figure*}[t]
		\centering
		\subfloat[RS (\# 147 of 1001)]{
			\includegraphics[width=0.7\columnwidth]{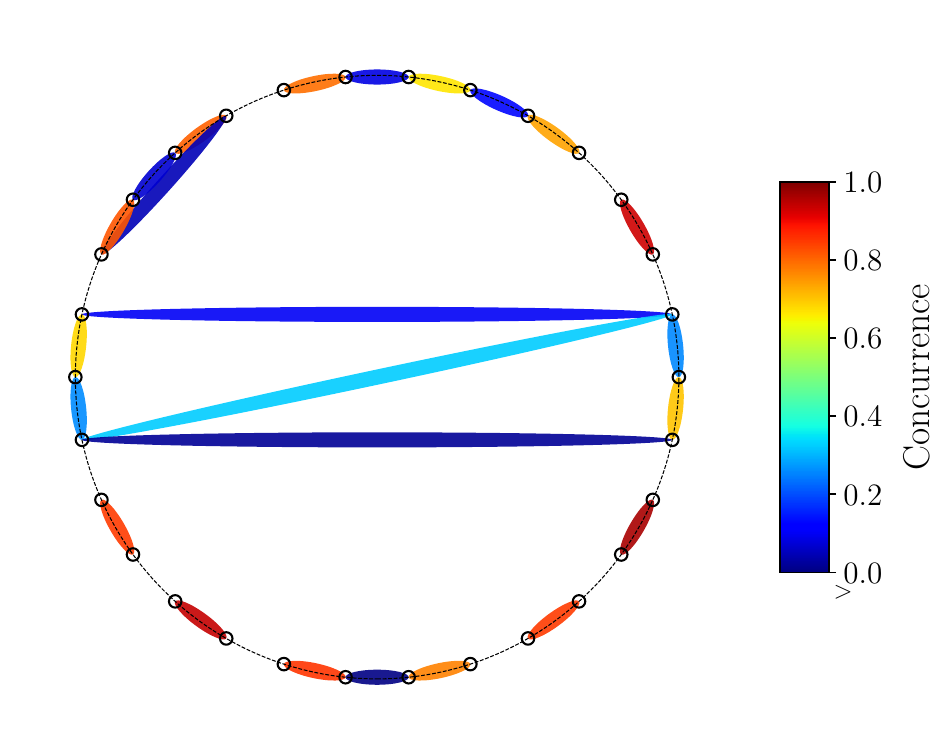}
			\label{fig:147disreal}}
		\subfloat[RS (\# 1000 of 1001)]{
			\includegraphics[width=0.7\columnwidth]{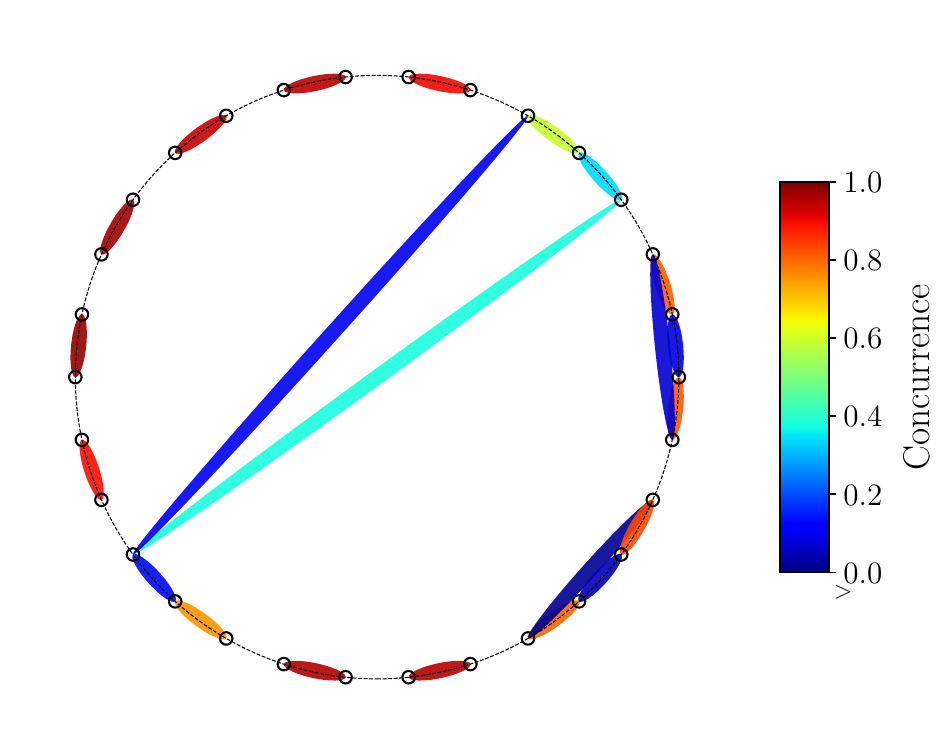}
			\label{fig:1000disreal}
		}
		\subfloat[TLL]{
			\includegraphics[width=0.7\columnwidth]{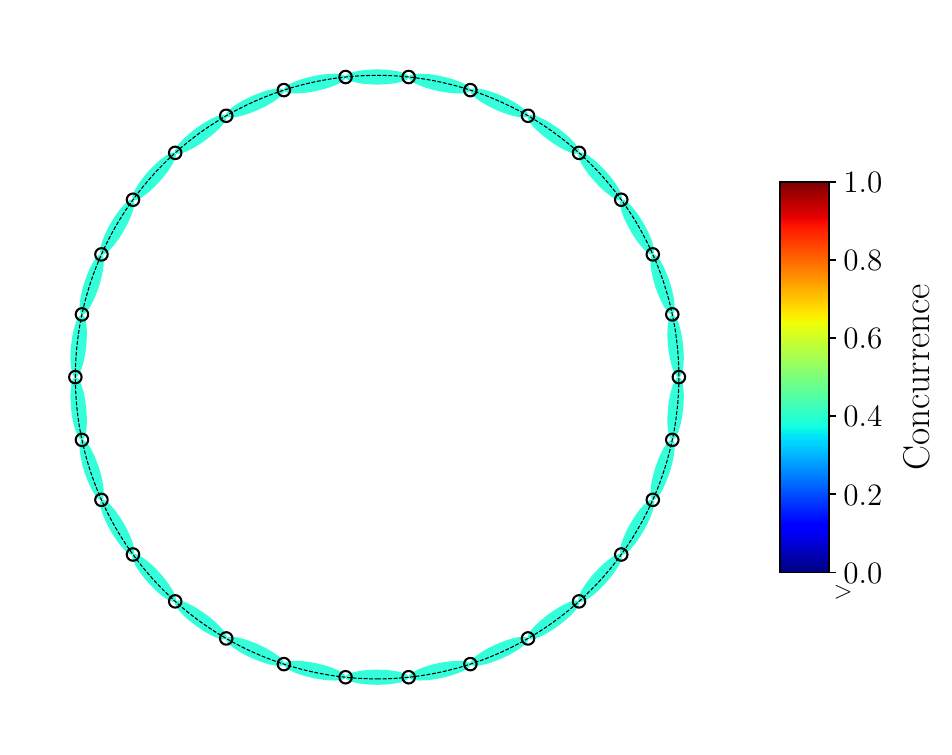}
			\label{fig:ConcdistTLL}
		}
		\caption{Visually demonstrating the difference in the concurrence distribution between the random singlet (RS) and Tomonaga Luttinger liquid (TLL) phase respectively. (a) Concurrence distribution of the $147^{\text{th}}$ disorder realization. (b) Concurrence distribution of the $1000^{\text{th}}$ disorder realization. (c) Concurrence distribution of the TLL. While TLL only has nearest neighbor concurrence, the RS phase has a non-zero concurrence for odd neighbors at all length scales. The plots here were obtained for a $N=30$ spin chain.}
		\label{DisorderRealiz}
	\end{figure*}

	As evident in Fig.~\ref{ConcT=0}, the TLL only has nearest neighbor pairwise entanglement  \cite{Gu2003a,Jin2004} while the RS also has a finite contribution coming from further neighbors as can be seen in Fig.~\ref{DisorderRealiz} for chosen disorder realizations. 
	\nic{
	The concurrence of the RS phase vanishes at even distance, $n = \{2, 4, 6, \ldots\}$,  
	while showing a slow decay at odd distance $n = \{1, 3, 5, \ldots\}$, consistent 
	with earlier calculations for disordered XY spin chains~\cite{Hoyos2006}.
	This echos the patterns found in spin correlations at long distances 
	where the decay is controlled by rare, strong singlet bonds \cite{Fisher1994,Hoyos2007}.
	None the less, it would be a mistake to identify the concurrence of the RS singlet phase 
	identically with its correlations, a point addressed in Appendix~\ref{appendix:worked.example}.
	}
	
	\nic{
	We note that, relative to the contribution from the nearest neighbors the concurrence 
	observed at longer length scales is quite small,} 
	and in fact from the point of view of INS probes, is not observable. 
	We explain this later in Sec.~\ref{OrderAvgsec}. 
	
	\subsection{TLL and RS are multipartite entangled}\label{sec:TLLandRSSMultipartite}
	\begin{figure}[h]
		\centering
		\includegraphics[width=1.\columnwidth]{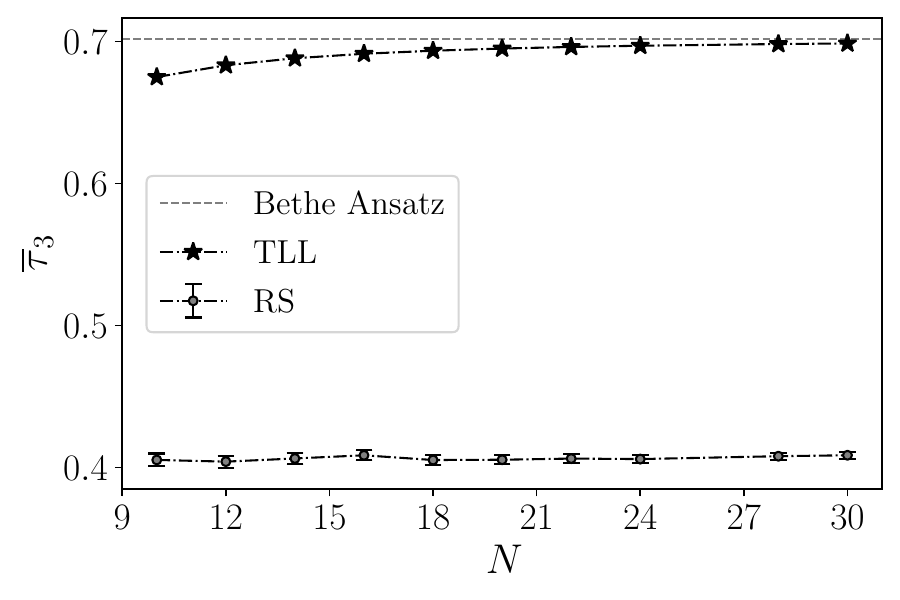}
		\caption{Finite residual tangle (RT) indicating the existence of multi-partite entanglement in the Tomonaga Luttinger liquid (TLL) and random singlet (RS) phase. RT of  TLL is approaching close to its Bethe Ansatz value as N increases. For the RS, RT remains fairly constant.}
		\label{RT}
	\end{figure}
	
	\begin{figure}[t]
		\vspace{-0.59cm}
		\centering
		\subfloat[QFI density vs $q$]{\includegraphics[width=1.08\columnwidth]{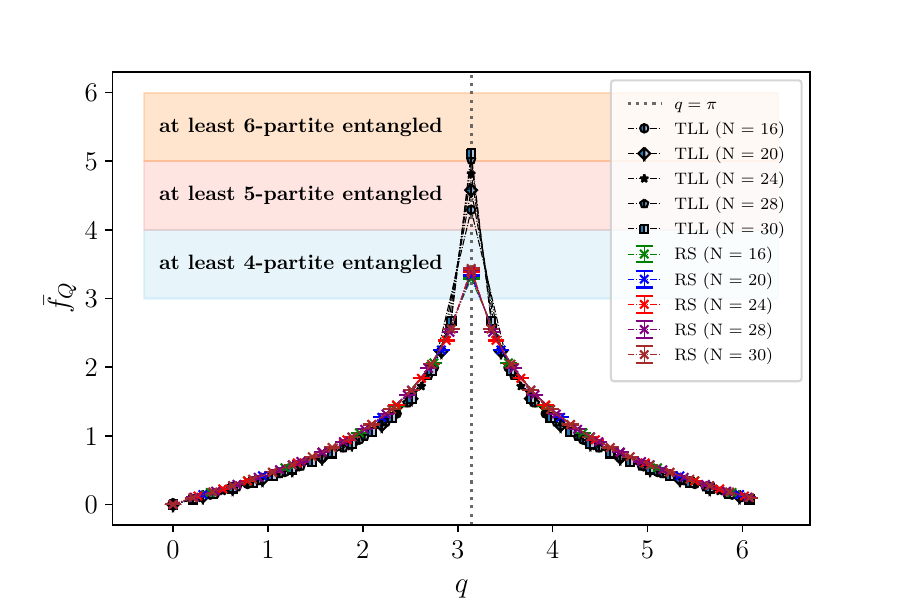}\label{fig:QFIvq}}
		\quad
		\subfloat[Scaling of QFI density at $q = \pi$]{\includegraphics[width=1.08\columnwidth]{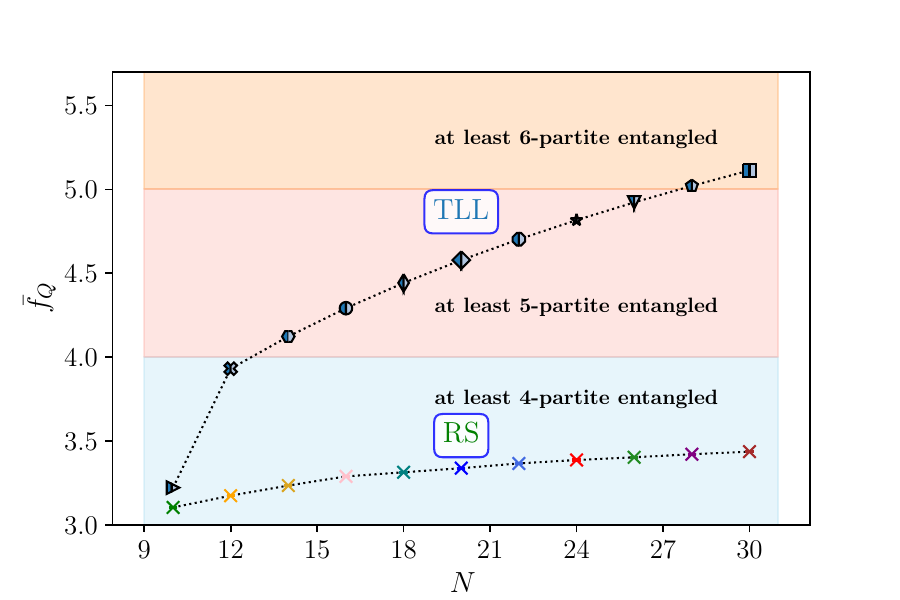}\label{fig:growentdepth}}
		\caption{Observing the growth of entanglement depth at the $q = \pi$ point in Tomonga Luttinger liquid (TLL) and  random singlet (RS) phase. (a) Quantum Fisher Information (QFI) density for TLL and RS phase plotted as function of $q$ for $N = 16, 20, 24,28,30$. (b) QFI density plotted as function of $N$ at $q = \pi$ point for TLL and RS phase. The rate of growth of entanglement depth is suppressed in the RS phase. Here we observe the TLL phase to be at-least $6$-partite entangled while the RS is at least $4$ partite entangled.}
		\label{fig:QFIT0}
	\end{figure}
	
	To further understand the entanglement structure of these phases next, we employ the multi-partite measures, i.e RT and QFI. We find that both the TLL and RS phase show a finite multi-partite entanglement as seen through the RT Fig.~\ref{RT} and QFI Fig.~\ref{fig:QFIT0}. 
	
	For the TLL, in Fig.~\ref{RT}, we observe the RT approaching
	\begin{equation}\label{RTTLLBA}
		\bar{\tau}_{3} = 0.701553\;,
	\end{equation}
	it's value in the thermodynamic limit, calculated using Bethe Ansatz~\cite{hulthen1938austauschproblem,Jin2004}.
	On the other hand, for the RS phase, the RT is fairly constant evidencing the existence of multi-partite entanglement.
	The multi-partite nature of the TLL \cite{Menon2023} and RS is also predicted by the QFI. At the $q = \pi$, we observe that the TLL is at-least 6-partite entangled and that the RS phase is at-least 4-partite entangled for the largest spin chain size ($N =30$) we access. From Fig.~\ref{fig:QFIT0} one can also see an increasing entanglement depth with system size both in the TLL and RS phase at the $q = \pi$ point. This growth, however, is more pronounced in the TLL as compared to the RS phase.  
	
	The implication of such a result is that if one observes a multi-partite entanglement using QFI in an experiment, then one cannot simply rule out the RS phase as one might have prematurely inferred from it's schematic picture. This is not in contradiction with the SDRG method. The fixed point of the SDRG method, the infinite randomness fixed point (IRFP), is a description of dominant physics at long length scales, whereas the multipartite entanglement observed here pertains to short length scales (discussed further in Sec~\ref{sec:Disc}). 
	The specific characteristics of this multi-partite entanglement will in general depend on the disorder distribution. 
	Nevertheless, this multi-partite information is encoded within the ground state of the disordered spin chain and has consquences when trying ruling out RS phase experimentally. 
	Whether there are \nic{further corrections to the structure of the RS state} \cite{Mohdeb2020}  at longer length scales is something we cannot answer from the system sizes we investigate.
	
	
	In the thermodynamic limit, one expects that the QFI for the TLL to diverge logarithmically \cite{Affleck1986}. On the other hand, for the RS phase, a complete suppression of QFI to a constant value is expected \cite{Hoyos2007,YuRong2016}.
		
	\nic{As a historical footnote, we remark that the presence of multipartite entanglement 
	in the RS state could have been anticipated from the arguments of Fisher, who envisaged 
	random singlets forming between effective \mbox{spin--1/2} objects of ``finite extent'' \cite{Fisher1994}.
	However at the time the SDRG method was introduced, ideas of entanglement yet to  
	permiate beyond the quantum information community.}
	
		
	\section{Finite Temperature results}\label{sec:FiniteTemp}
	Having discussed the zero temperature entanglement structure of the two phases, we move to explore the effects of finite temperature on this structure. For the task at hand, we employ concurrence and the equal-time structure factor as entanglement witness based measures. In Sec.~\ref{OrderAvgsec}, we discuss the importance of realizing what is being experimentally accessed and how it affects the way one performs the disorder-average in their analysis.
	In particular, we highlight the importance of knowing how to perform the disorder-average keeping in mind the observables accessible in an experiment. We point that this is especially crucial when one is trying to rule out disorder induced phases by using certain entanglement measures. Following which, in Sec.~\ref{LTgrowthEW}, we find that the low-temperature growth of entanglement witness based measure turns out to be very useful to characterize these phases. 
	
	\subsection{The order of disorder-average matters for concurrence}\label{OrderAvgsec}
	If one goes by our previous results for concurrence then one expects to see beyond nearest neighbor 2-partite entanglement in the RS phase.  Does this imply that if experimentally we do not observe further nearest neighbor concurrence can we rule out the presence of a disorder induced phase? On the practical side, can the further nearest neighbor entanglement even be observed in INS? Unfortunately, the answer to both these questions turns out to be negative. Typically, in experimental situations such as in INS, we access disorder-averaged spin correlations. What this means is that when characterizing entanglement from spin correlation for example, one does not access the entanglement per disorder realization. While this distinction might feel academic, for measures like concurrence, the order of how one performs to disorder-average becomes important if one is comparing a theoretical model against the experimental predictions. \\
	
\begin{figure}[t]
		\centering
		\subfloat[Disorder-averaged Concurrence at finite temperature]{\includegraphics[width=1.\columnwidth]{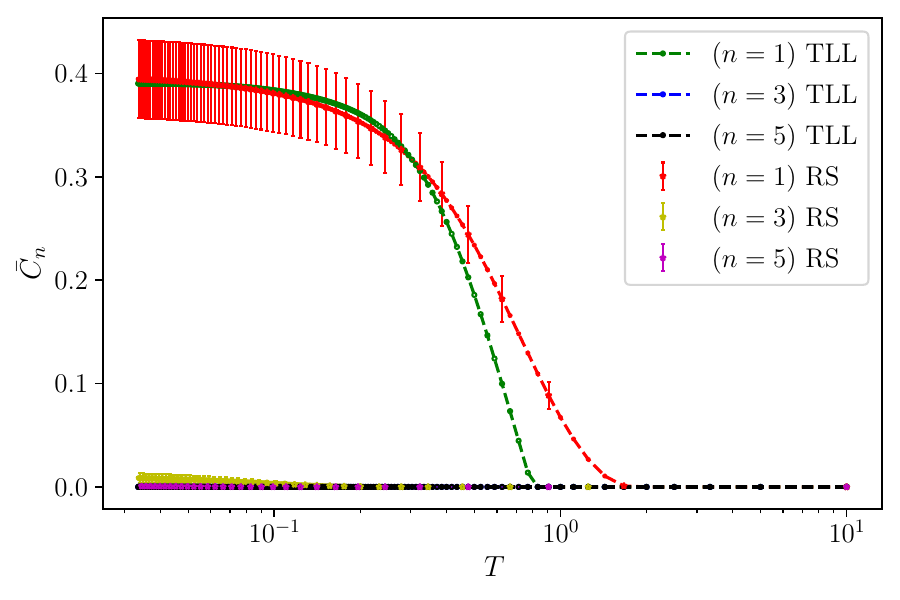}\label{fig:DisAvgConc}}
		\quad
		\subfloat[Concurrence from disorder-averaged spin correlations]{\includegraphics[width=1.\columnwidth]{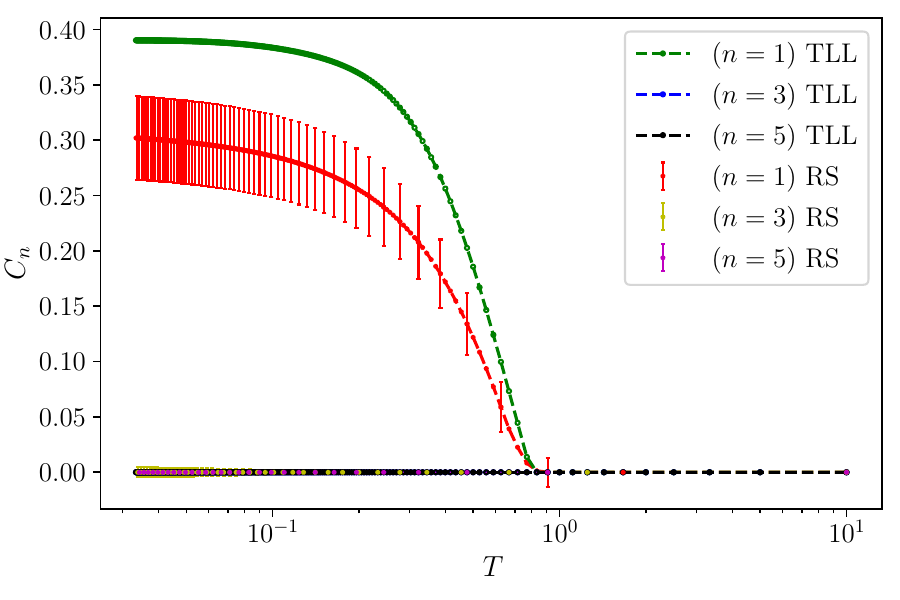}\label{fig:ConcDisAvgCorr}}
		\caption{Demonstrating how the order of disorder-average affects the concurrence distribution in the random singlet (RS) phase. (a) Disorder-averaged concurrence observed for $1$st, $3$rd and $5$th neighbors as function of temperature for $N = 20$. (b) Concurrence, obtained from disorder-averaged spin correlations, observed for the $1$st, $3$rd and $5$th neighbors as a function of temperature. For (a), concurrence is evaluated using Eq.~(\ref{concfromspin}) for every disorder realization and then averaged. On the other hand, for (b), concurrence is evaluated using Eq.~(\ref{concfromspindisordavg}). in which case the pairwise entanglement distribution of the RS phase mimics the Tomonga Luttinger liquid (TLL) phase i.e., becomes nearest neighbored. 
			To help read the plots some of the error bars coming from disorder averages have been suppressed.}
		\label{ConcfinT}
	\end{figure}

	To begin with, let's look at the finite temperature behavior of concurrence with the way we have been evaluating it as explained in Sec.~\ref{ConcT=0RSvTLL}. In Fig.~\ref{ConcfinT}, we find that up to some threshold temperatures the RS phase has finite pairwise entanglement from further neighbors whereas the TLL phase has only nearest neighbor pairwise contribution consistent with what we saw before. 
	
	On the other hand, we can also evaluate the disorder-average of the spin correlations first and then evaluate the concurrence as below
	\begin{equation}\label{concfromspindisordavg}
		C_{ij} = 2\max\bigg\{0,  2 |\bar{g}^{zz}_{ij}| - \bigg|\frac{1}{4}+\bar{g}^{zz}_{ij}\bigg|\bigg\}\;,
	\end{equation}
	where $\bar{g}^{zz}_{ij}$ is the disorder-averaged spin correlation function evaluated as per Eq.~(\ref{disavg}).
	From the experimental perspective, it is this concurrence that would be accessed. In Fig.~\ref{ConcfinT}, we perform this and find that RS has only nearest neighbor 2-partite entanglement just like the TLL. We do not observe any finite beyond nearest neighbor concurrence as one might have expected. What this means is that experimentally one would not be able to observe beyond nearest neighbor pairwise entanglement if they access disorder-averaged spin correlations. Thus ruling out the RS phase by looking at the 2-tangle or concurrence and observing no beyond nearest neighbor 2-partite entanglement is incorrect. We also observe in Fig.~\ref{fig:ConcDisAvgCorr} that the nearest neighbor concurrence vanishes at the same temperature for both the TLL and RS phase.
	
	\begin{figure}
		\centering
		\includegraphics[width=\columnwidth]{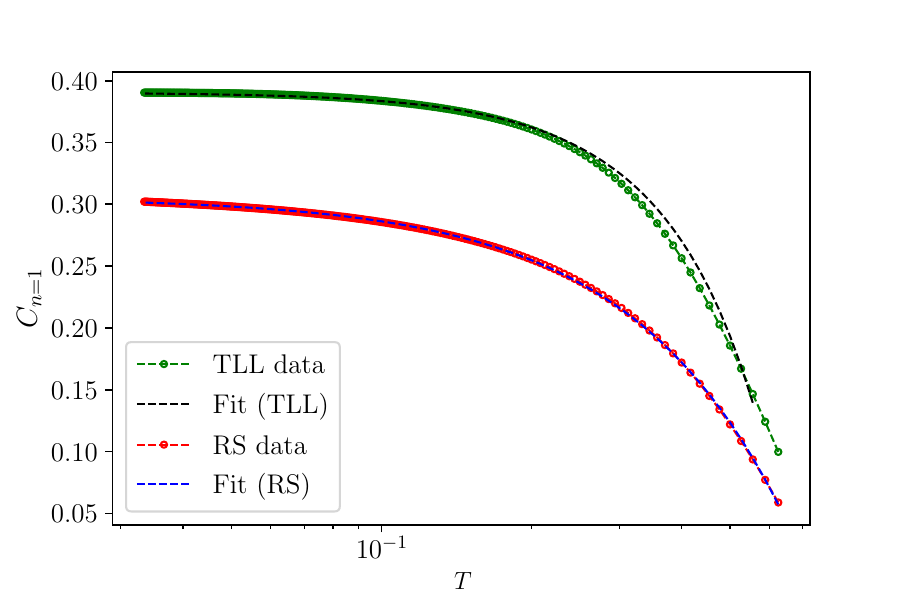}
		\caption{Extracting central charge information  from the low-temperature behavior of concurrence for the Tomonga Luttinger liquid (TLL) and random singlet (RS) phase respectively. The plots here are evaluated for $N = 20$ spin chain. For the TLL phase, we make use of Eq.~(\ref{eq:ConcfromInternal}) and Eq.~(\ref{eq:InternalTLL})  to fit the observed low-temperature behavior. For the RS phase, we make use of Eq.~(\ref{eq:ConcfromInternal}) and  Eq.~(\ref{eq:InternalRS}) for the observed low-temperature behavior. Note that here we are evaluating the concurrence from disorder-averaged spin correlations [Eq.~(\ref{concfromspindisordavg})] as one would access in experiments like inelastic Neutron scattering (INS). For ease of viewing, the error bars have been suppressed.}
		\label{ExptConc_Central}
	\end{figure}
	
	\subsection{The low-temperature behavior of concurrence contains central charge}
	The low-energy behavior of TLL can be described by a conformal field theory with a central charge, $c_{\sf TLL} = 1$ \cite{Affleck1986}. By noting that the nearest neighbor concurrence for a TLL can be related to internal energy, one can easily obtain the central charge as follows. The key is to realize that  the nearest neighbor concurrence can be related to the internal energy as follows.
	\begin{equation}\label{eq:ConcfromInternal}
		C_{1} = \max\{0,2 |u(T)| - 0.5\}\;,
	\end{equation}
	where 
	\begin{equation}\label{eq:InternalTLL}
		u(T) = u(0) + c T^2\bigg( \frac{1}{3} + \frac{3+\log(\pi/T)}{8\log(\pi/T)^4}\bigg)\;,
	\end{equation}
	is the low-temperature behavior of the internal energy \cite{Klumper1998} and $c$ is the central charge.  
	
	We put this to test in Fig.~\ref{ExptConc_Central} and find a good agreement with the our fit 
	\begin{equation}
		u(0) \approx -0.445\;,\; c_{\sf TLL} \approx  1.01
	\end{equation}
	On the other hand, it has been shown that the RS phase is characterized by an effective central charge \cite{Rafael2004,Laflorencie2005,Hoyos2007,Mohdeb2020}
	\begin{equation}
		c_{\sf RS} = c_{\sf TLL}\ln{2}\;.
	\end{equation}
	We conjecture that it is possible to access this information from the concurrence of the RS phase where we employ the disorder-average that is experimentally relevant. 
	We find the following ansatz fits very well with the observed low-temperature behavior for the RS phase
	\begin{equation}\label{eq:InternalRS}
		u(T) = u(0) + \frac{c_{\sf eff}T^d}{3}\;,
	\end{equation}
	where $d$ is a fit parameter and
	\begin{equation}
		c_{\sf eff} = c\log(2)\;,
	\end{equation} is what we interpret as the effective central charge. 
	In Fig.~\ref{ExptConc_Central} we obtain 
	\begin{equation}
		u(0) \approx  -0.403\;,\; c_{\sf eff} \approx 0.71 \;,\; d \approx 1.39 \;,
	\end{equation}
	where the parameter $u(0)$ is size-dependent, and we find $c$ to be very close to $c_{\sf TLL}$. 
	
	\subsection{
		The low-temperature growth of multi-partite entanglement is a useful probe }\label{LTgrowthEW}
\begin{figure*}[t]
		\centering
		\subfloat[]{\includegraphics[width=1\columnwidth]{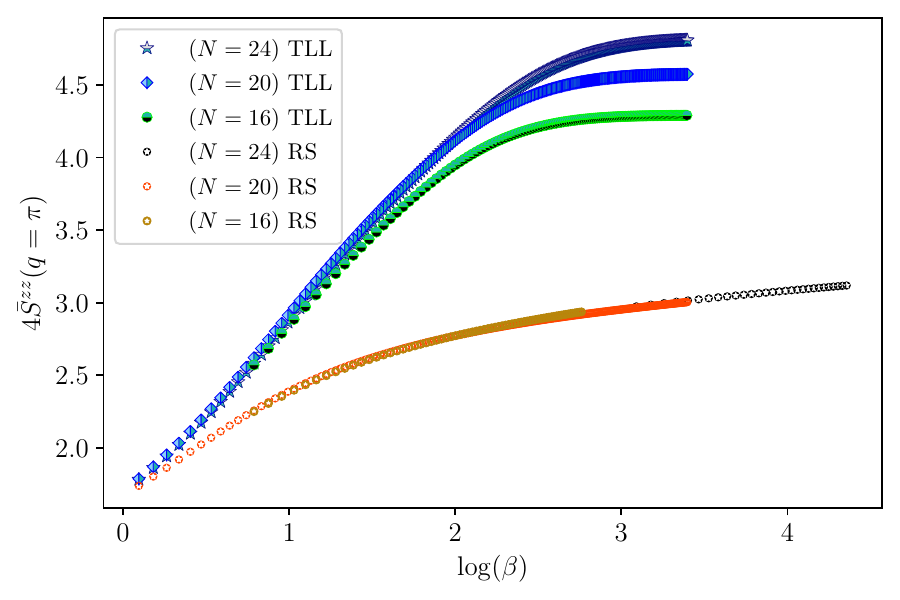}\label{fig:SqTforTLLandRS}}
		\subfloat[]{\includegraphics[width=1\columnwidth]{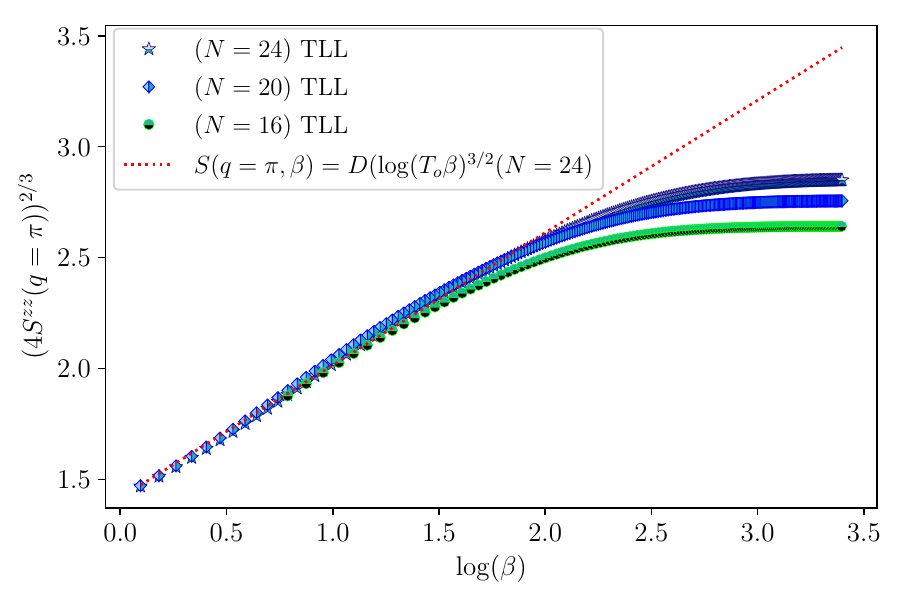}\label{fig:SqTforTLL}}
		\quad
		\subfloat[]{\includegraphics[width=1\columnwidth]{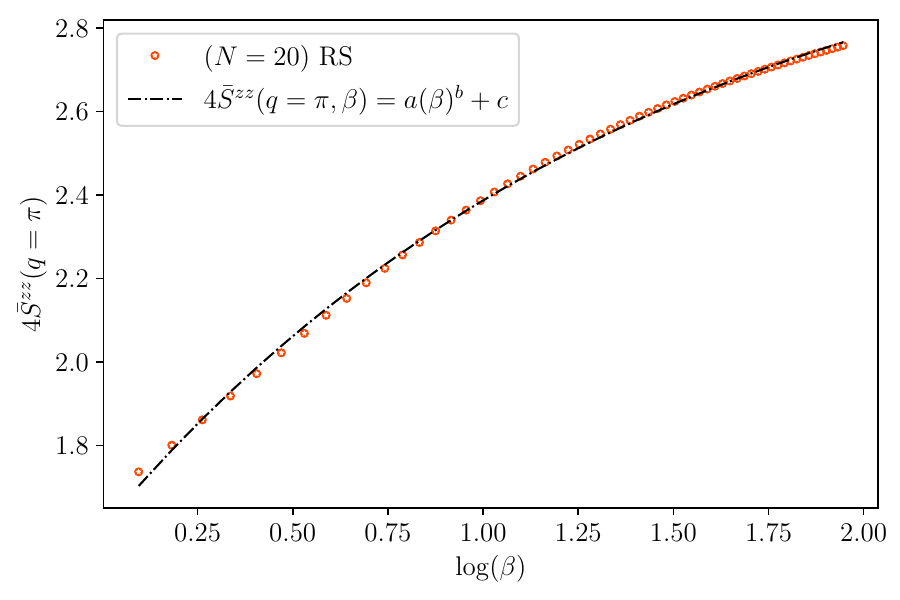}\label{fig:SqTforRS}}
		\caption{Comparison of the low-temperature behavior of the equal time structure factor for the Tomonga Luttinger liquid (TLL) and random singlet (RS) phase. (a) Equal time structure factor of both the TLL and RS phases observed as function of $\log(\beta)$ for $N =16,20,24$ chains. (b) Testing the low-temperature behavior of equal time structure factor for the TLL against the analytic prediction Eq.~(\ref{eq:SqTTLL}). (c) Testing the  low-temperature behavior of equal time structure factor for the RS against the ansatz Eq.~(\ref{eq:SqTRS}) for $N =20$ spin chain. The fitting parameters are given in Eq.~(\ref{eq:fitSqTLL}) and Eq.~(\ref{eq:fitSqRS}). The fits were checked within a fixed interval $0.09<\log(\beta)<1.9$ for both the TLL and RS phases. This choice was made owing to the finite size effects in the TLL phase for temperatures below the lower bound. For ease of viewing, the error bars have been suppressed.}
		\label{equalstructfactor}
	\end{figure*}

	We saw that the zero temperature QFI, Fig.~\ref{fig:QFIT0}, predicts both the TLL and RS phase to be multi-partite entangled. 
	Thus, stating whether a phase is $m$-partite or some other is $l$-partite is not a very useful probe if we want to distinguish them experimentally. 
	Instead, we propose  to investigate the low-temperature behavior of multi-partite entanglement witness measures. We employ the equal time structure factor for this task. As evident in Figs.~\ref{equalstructfactor}, the growth of entanglement at low-temperatures is qualitatively quite different. For the TLL, the low-temperature scaling behavior is well known \cite{Starykh1997}
	\begin{equation}\label{eq:SqTTLL}
		4S^{zz}(q = \pi,\beta) =  D \log(T_{o}\beta)^{3/2}\;.
	\end{equation}
	We confirm that Fig.~\ref{fig:SqTforTLL}
	\begin{equation}\label{eq:fitSqTLL}
		D \approx 0.46\;,\; T_{o} \approx 10\;,
	\end{equation}
	up to a range of temperatures our results match very well with Eq.~(\ref{eq:SqTTLL}), however at lower temperatures, we do observe finite size effects for the different spin chain sizes we explore. 
	
	On the other hand, we find the growth of entanglement to be quite slow in our disorder induced phase. We propose the following empirical behavior,
	\begin{equation}\label{eq:SqTRS}
		4S^{zz}(q=\pi,\beta) = a \beta^{b} + d \;,
	\end{equation}
	which we found for 
	\begin{equation}\label{eq:fitSqRS}
		a \approx  -1.5 \;,\; b \approx -0.6\;,\; d \approx 3.1
	\end{equation}	
	to describe our data Fig.~\ref{fig:SqTforRS} very nicely. These parameter values are system size and disorder strength of disorder dependent, however. 
	
	\section{Discussion}\label{sec:Disc}
	\subsection{ Multi-partite entanglement localization and understanding RS `phase' }
	
\begin{figure*}[t]
		\centering
		\subfloat[]{\includegraphics[width=1.6\columnwidth]{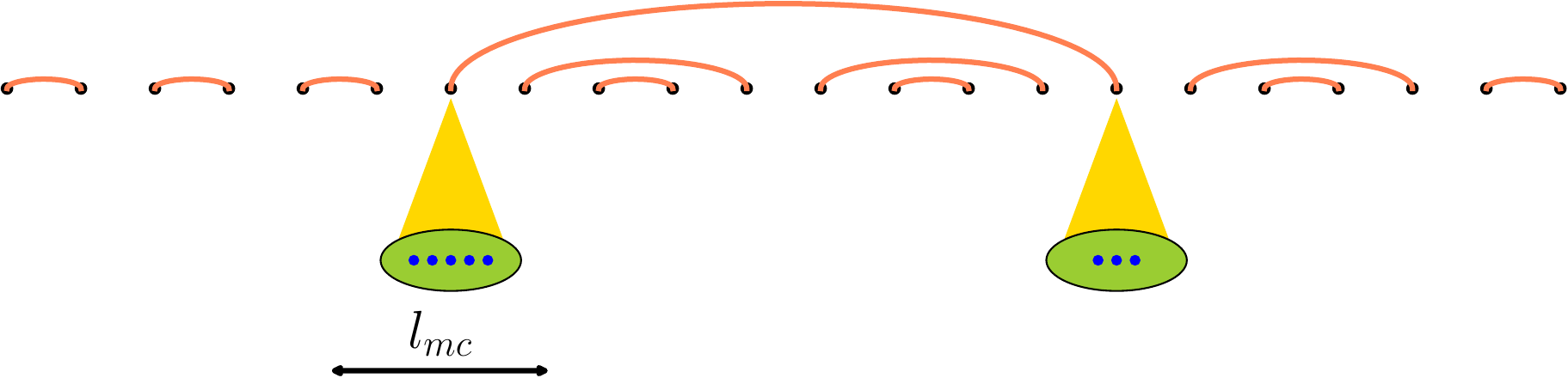}\label{fig:SDRGRSstatecarricature}}
		\quad
		\subfloat[]{\includegraphics[width=1.6\columnwidth]{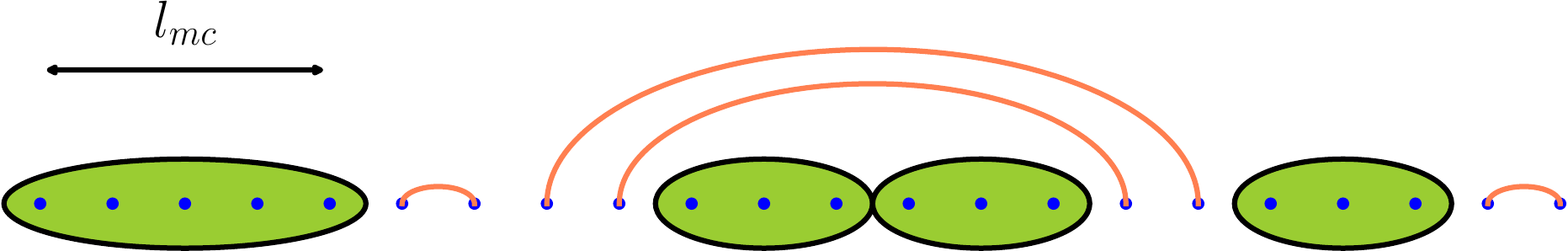}	\label{fig:RSstatecarricature}}
		\caption{An illustration of the random singlet (RS) phase. (a) The schematic picture of the RS phase, as described within the strong disorder renormalization group (SDRG), where pairs of entangled effective spins (black) are connected by orange colored arcs. The green shaded ovals highlight that the effective spins are made up of multiple real space spins.  (b) Schematic picture of the RS phase predicted from our results, where multi-partite clusters (mc) are represented by green shaded ovals of size $l_{mc}$ alongside pairwise entangled sites (blue) connected by orange colored arcs. These mc are localized to distances below the crossover length $L_c$.}
	\end{figure*}

	Within the SDRG approach, the RS phase is pictured as a tensor product of singlets formed out of effective spins per disorder realization Fig.~\ref{fig:SDRGRSstatecarricature} \cite{Fisher1994}. On the other hand, both RT Fig.~\ref{RT} and QFI  Fig.~\ref{fig:QFIT0}, point to the RS phase to be multi-partite entangled.  Is there an apparent contradiction here? No, there isn't. 
	The key here is to realize that the singlets in the SDRG picture is between effective spins \cite{Fisher1994}. 
	The short range physics is scaled away in the renormalization procedure and it is this short length scale physics that contributes to our observation of multi-partite entanglement.

	The SDRG framework provides an extremely good description of the long-wavelength behavior of a disordered spin chain~\cite{Laflorencie2004, Laflorencie2005}.
	%
	%
	However this long-wavelength picture, in which correlations are controlled by an infinite--disorder fixed point, should 
	be understood to come into play above a crossover length, $L_c$ \cite{Laflorencie2004}. 
	This crossover can be observed in Fig.~\ref{fig:QFIvq}, as the momentum 
	\begin{equation}\label{eq:crossmomen}
		q_c = \pi - \frac{2 \pi}{L_{c}} \;,
	\end{equation}
	below which the RS phase and the TLL are practically indistinguishable. What this means is that at distances shorter than $L_c$, spins can be entangled in clusters that contribute to the  observed multipartite entanglement, 
	as illustrated in Fig.~\ref{fig:RSstatecarricature}. 
	However for distances $ \gg L_c$, these `multi-partite' clusters are completely suppressed,
	where now the dominant entanglement is at most 2-partite. 
	%
	At short length scales, the multi-partite entanglement plays a role, as witnessed by  the QFI density [Fig.~\ref{fig:QFIvq}].  
	We understand this as the \textit{localization of the multi-partite entanglement}. 
	
	We can turn this observation into a bound on QFI density.
	If the RS phase, as understood within the SDRG, is at most 2-partite entangled, 
	this places a limit on the entanglement contributed by long length-scales.  
	We therefore expect
	\begin{equation}\label{eq:QFIinequal}
		\bar{f}_{Q}(q = \pi) - \bar{f}_{Q}(q = q_c) \leq 2 \;,
	\end{equation}
	A violation of this inequality would rule out the RS phase. \\
	
	To summarize the above discussion: 
	the RS phase contains multi-partite entanglement at short length scales. 
	This is not in contradiction with the SDRG approach, where multi-partite entanglement information is 
	absorbed within effective spin-1/2 moments, which go on to form singlets at long length scales.
	And as we have documented, the multi-partite entanglement found in numerical simulations 
	of disordered spin chains originates in contributions below a crossover scale i.e., from short--range physics. 
	In short, while multi-partite entanglement is not relevant to correlations at the longest length scales, 
	which are controlled by the SDRG fixed point, it does exist.   
	As a consequence, the observation of finite multi-partite entanglement using QFI 
	cannot be used to rule out a RS phase.
	
	\subsection{Infinite disorder and Thermodynamic limit.}
	Up until now we focussed on a setting where we considered a box distribution as specified in Eq.~(\ref{eq:BoxDist}) with $\Delta = 1.0$. How does the understanding change for some other distribution? To make this question more concrete let us consider the following family of distributions
	\begin{equation}
		P(J;\delta)= \begin{cases}A \delta^{-1}J^{\delta^{-1} -1} & \text{for } 0 \leq J < J_{max} \\ 0 & \text{otherwise}\;, \end{cases}
	\end{equation}
	where 
	\begin{equation}
		A = 1/J_{max}^{1/\delta} \;.
	\end{equation}
	Note that for $\delta = 1$ and $J_{max} = 2$ we recover the boxed distribution Eq.~(\ref{eq:BoxDist}). In the limit $\delta \to \infty$, one obtains the infinite randomness fixed point (IRFP) of the SDRG method. 
	
	\begin{figure*}[t]
		\centering
		\subfloat[QFI density for different $\delta$]{\includegraphics[width=1\columnwidth]{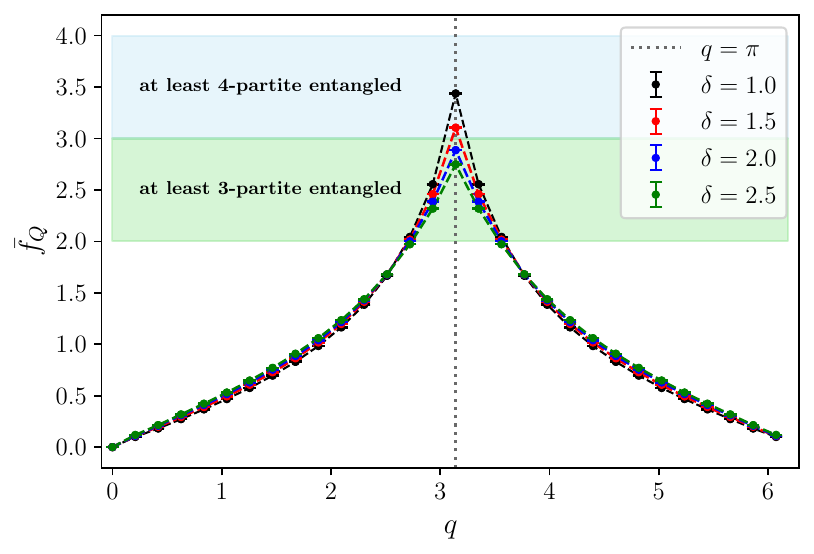}\label{fig:QFIdelta}}
		\subfloat[RT for different $\delta$]{\includegraphics[width=0.96\columnwidth]{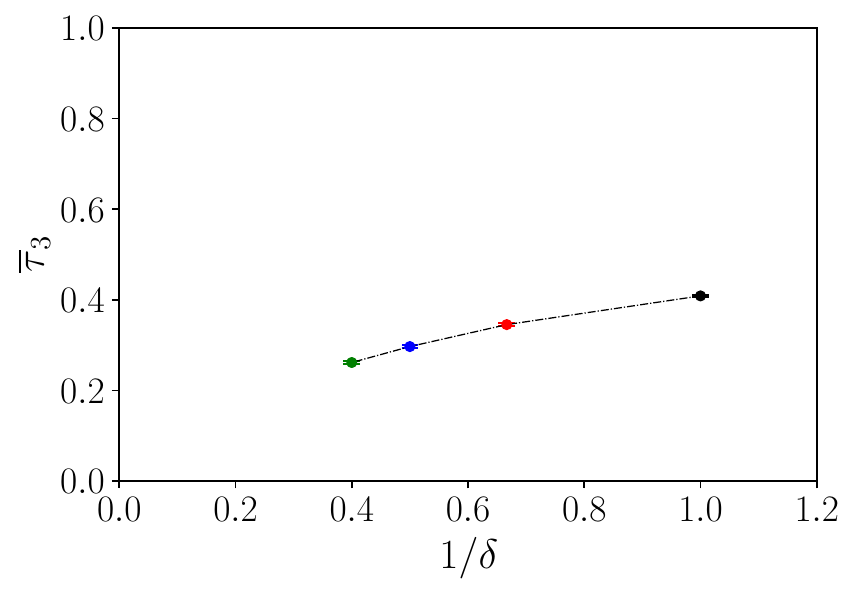}\label{fig:RTdelta}} 
		\caption{Observing the decreasing entanglement depth with increasing $\delta$ in the random singlet (RS) phase. (a) Quantum Fisher information (QFI) density as function of q for $\delta = 1.0,1.5,2.0,2.5$ in a $N = 30$ spin chain. (b) Residual tangle (RT) as a function of $\frac{1}{\delta}$ for $N = 30$ chain, where the line is used as a guide. While an increasing $\delta$ reduces the entanglement depth, for $\delta = 2.5$ and $N=30$ spin chain the QFI density at the $q = \pi$ still detects a multi-partite entangled RS phase. }
		\label{fig:DeltaQFIandRT}
	\end{figure*}
	For starters, an increasing $\delta$ reduces the observed entanglement depth at the $q = \pi$ point as seen in Fig.~\ref{fig:DeltaQFIandRT}. This in turn moves the crossover momentum $q_c$ Eq.~(\ref{eq:crossmomen}) further away from $\pi$. Thus, Eq.~(\ref{eq:QFIinequal}) remains valid for increasing $\delta$ as well. The disorder parameter $\delta$ controls the crossover length and which consequently bounds the maximum length of the multi-partite cluster, $l_{mc}$ 
	\begin{equation}
		l_{mc} \leq L_c \;,
	\end{equation}
	such that the role of these clusters become increasingly negligible above the crossover length.
	What does one expect in the thermodynamic limit? At least, for a finite $\delta$, say below $\delta \leq \delta^{*} = 2.5$, from our results, we do expect to observe an entanglement depth greater than $2$
	\begin{equation}
		\lim_{\delta \to \delta^{*}}\lim_{N \to \infty} f_{Q}(q = \pi) = C_{*} > 2 \;,
	\end{equation}
	where $C^*$ is some constant. 
	For real materials, a finite $\delta$ would already imply an entanglement depth larger than two.
	For instance, in \cite{Volkov2020}, it was reported that Ba$_5$CuIr$_{3}$O$_{12}$ single crystals can characterize a RS phase with a $\delta \approx 2.5$. 
	
	\subsection{Experimentally ruling out RS phase is a  subtle business}
	The RS phase is multi-partite entangled. This could have been inferred from the results in \cite{Hoyos2007,YuRong2018}, in hindsight.
	While the 2-partite entanglement of this phase may indeed be the relevant entanglement setting to understand the effects of disorder, it does also have a multi-partite contribution from length scales below the crossover length. Even though the multi-partite nature of the phase may not  be relevant to understand the effects of disorder at long length scales, these contributions do not simply disappear either. Consequently, ruling out RS phase under the assumption that it has only 2-partite entanglement is incorrect. 
	Furthermore, we showed how one might be misled into ruling out RS phase by the absence of any concurrence beyond the nearest neighbors. One might incorrectly conclude the absence of long range singlets from such an observation. 
	
	Convex roof measures like concurrence depend quite a lot on the way the disorder-average is performed.  This is due to the presence of the `$\max$' function in the definition Eq.~(\ref{conc}). The operation of averaging over disorder realizations does not commute with the max operation.  Whether one takes the disorder-average inside the max function, or outside it, the end results are completely different. Why is this important? For quenched disordered systems, the spin correlations are already disorder-averaged. Thus if one were to access spin correlations from experiments like INS and from those tried evaluated concurrence using  Eq.~(\ref{concfromspindisordavg}) then one would observe  Fig.~\ref{fig:ConcDisAvgCorr}  as opposed to Fig.~\ref{fig:DisAvgConc}. The former corresponds to the case when concurrence has been evaluated from disorder-averaged spin correlations and the latter is the one where we evaluate concurrence for each disorder realization and then average it. In the former case, the RS phase shows no further neighbor concurrence, essentially mimicking the TLL phase at the 2-partite level Fig.~\ref{fig:ConcdistTLL}. 
	In \cite{Scheie2023}, the authors incorrectly ruled out RS phase by observing vanishing concurrence and multi-partite entanglement using QFI. 
	However, both of these observations cannot rule out the RS phase as shown in Sec.~\ref{sec:TLLandRSSMultipartite} and \ref{OrderAvgsec}. 
	This point will be discussed further in a follow-up work \cite{Tokuro2024}.
	\subsection{What can be useful experimentally?}
	For 1d critical systems, the low-temperature behavior of concurrence can access central charge information. In particular, the nearest neighbor concurrence of an $SU(2)$ symmetric phase can be related to the internal energy, and this allowed us to extract the central charge of the TLL. On the other hand, for RS phase, while the order of disorder-average effects how concurrence is distributed among the sites, the universal information about the phase should still be accessible if it is contained within the 2-partite entanglement. This universal information is the effective central charge. Since experimentally, the concurrence distribution for the RS phase can be nearest-neighbored just as the TLL, we thought that perhaps one might be able to access the effective central charge for the RS phase. This is based on the assumption that there exists a form of generalized c-theorem for the RS phase~\cite{Rafael2004}. This is what we put to test in Fig.~\ref{ExptConc_Central} and see some positive indications. However, we admit that this is conjectural at best.\\
	
	On the other hand, multi-partite entanglement witnesses like the QFI or the equal time structure factor are invaluable markers for not only identifying the existence of multi-partite entanglement but also how disorder affects their growth.  We saw this by the marked difference in the low-temperature behavior of the equal time structure factor in the TLL and RS phase. This behavior can be experimentally useful in identifying the role of disorder in spin chains.\\

	\subsection{Limitations to detection of multi-partite entanglement}	
	Measures like QFI, when accessed via the dynamic structure factor, can only detect multi-partite entanglement if the underlying phase shows correlations peaked around specific $\mathbf{q}$ points. Due to the sum rule, this structure constraints the visibility of multi-partite entanglement detection at other $\mathbf{q}$ points. However, such peaked structure factors are not universal. For instance, in Kitaev and Kagome QSLs, there are no special $\mathbf{q}$ points with sharp features \cite{laeuchli2009dynamical, Knolle2014, Shimokawa2016}. Therefore, multi-partite entanglement cannot be accessed using INS in these cases. This does not imply the absence of multi-partite entanglement in these phases, rather, it suggests that the multi-partite entanglement could be spread among non-local correlations, rendering the operator used to access this information insufficient.
	
	\section{Conclusions}\label{sec:Conclusion}
	While there are no perfectly clean samples in nature, none are infinitely disordered either. In this work, we set out to understand how one can experimentally access the effects of disorder on the entanglement of ground states within condensed matter systems. We utilized the quantum spin chain as a representative example to investigate the impact of bond disorder. In this setting, we focussed on distinguishing between Tomonaga-Luttinger liquid (TLL) and random singlet (RS) phases through experimentally accessible entanglement measures, like concurrence, residual-tangle (RT) and quantum fisher information (QFI). \\
	
	We saw that both the TLL and RS phases are multi-partite entangled. For the TLL, the observed multi-partite entanglement is a 
	consequence of the uninhibited growth of correlations. On the other hand, such correlations are completely suppressed for the RS phase. Consequently, the observation of multi-partite entanglement in the RS phase is consequence of the localization of multi-partite clusters up to a crossover length-scale. 
	Above this regime, the relevant entanglement might indeed be 2-partite, consistent within the framework of strong disorder renormalization group (SDRG). To better characterize the effect of disorder, we introduced the inequality Eq.~(\ref{eq:QFIinequal}) which we found is also in agreement for the larger system sizes that were accessed in \cite{Hoyos2007, YuRong2018}.\\
	
	For quenched disorder, the order of average matters for convex roof measures. In particular, the distribution of pairwise entanglement, as characterized by concurrence, depends on how one performs the average. For the RS phase, this distribution can transform from being shared between distant sites Fig.~\ref{fig:DisAvgConc} to only among nearest neighbors Fig.~\ref{fig:ConcDisAvgCorr}, as is the case in TLL. Since in INS one would access disorder-averaged correlations, it is the latter that one would observe rather than the former. \\
	
	For the TLL, it is well established that it can be described by a conformal field theory. We leveraged this connection to extract the associated central charge from the low-temperature behavior of concurrence. Conversely, for the RS phase, we extracted the effective central charge from the low-temperature behavior of concurrence by employing disorder averaging accessible in experiments. This relies on appreciating that the order of disorder averaging alters the distribution of pairwise entanglement while preserving the universal information encoded within it. However, the existence of an effective central charge for the RS phase remains conjectural, based on the assumption of a generalized c-theorem \cite{Rafael2004}.\\
	
	From the experimental point of view, a more robust characteristic is to examine the low-temperature behavior of multi-partite entanglement witnesses like quantum Fisher information (QFI) or the equal time structure factor. We made use of  the equal time structure-factor to show a marked difference in the rate of growth of multi-partite entanglement within the two phases  Fig.~\ref{fig:SqTforTLLandRS}. This scaling provides a clear distinction between the TLL and RS phases, which can be observed experimentally.\\
	
	In the end, we would like to conclude, by pointing out that, in general some caution should be maintained especially when ruling out disorder-induced phases using measures based on witnesses. However, when used carefully, these can provide important information about quantum spin systems in the presence of disorder.
	
	\section*{Acknowledgements}
	This work is supported by the Theory of Quantum Matter Unit of the Okinawa Institute of Science and Technology Graduate University (OIST).	We would like to thank Allen Scheie,  Alan Tennant, and Francis Pratt for sparking our interest in this problem. We also thank Matthias Gohlke, Jos\'{e} Hoyos, Pranay Patil, Geet Rakala, and Anders Sandvik for their helpful suggestions, discussions and useful comments on this work. The calculations done here were performed using H$\Phi$ \cite{KAWAMURA2017}. This study is also supported by JSPS KAKENHI Grant Numbers 21K03477, 22H05266, 24H00974 and 25K07213. Majority of the numerical calculations were performed using the Core Facilities of the  Scientific Computing and Data Analysis section at OIST. Some parts were also done using the
	facilities of the Supercomputing Center, at ISSP, University of Tokyo.
	
\begin{appendix}
	
\section{Concurrence within pure and mixed states}
\label{appendix:worked.example}

	
	In this Appendix we consider two different ways of evaluating concurrence for a pair of qubits, one appropriate 
	for a mixed state, and one for pure state, with the aim of highlighting the difference between the two.
	
	For a \textit{mixed state} describing a pair of spins (qubits), concurrence is given by~\cite{Hill1997,Wootters1998}, 
	\begin{equation}\label{eq:concmixed}
		C_{ij}^{\sf mix} \equiv C(\rho^{\sf mix}_{ij}) = \max\{0, \lambda_{1} - \lambda_{2} -\lambda_{3}-\lambda_{4}  \} \; .
	\end{equation}
	where $\lambda_\nu$ are the solutions of 
	\begin{eqnarray}
		\rho_{ij} \tilde{\rho}_{ij} \ket{v_{\lambda_\nu}} &=& \lambda_\nu^{2} \ket{v_{\lambda_\nu}} \;,  
	\end{eqnarray}
	arranged in descending order, with 
	\begin{eqnarray}		
		\tilde{\rho}_{ij} &=& (\sigma^y \otimes \sigma^y) \rho^*_{ij}  (\sigma^y \otimes \sigma^y)\;,
	\end{eqnarray}
	where $^*$ reflects complex conjugation in a computational basis defined 
	in terms of eigenstates of $\sigma^z$ 
	(cf.~Eq.~(\ref{conc}), Eq.~(\ref{def:lambda}) and Eq.~(\ref{def:matrix}) 
	of Section~\ref{sec:EntanglBackgrnd}).
	
	In contrast, if the (reduced) density matrix describing the pair of spins reflects 
	a \textit{pure state}, then one can evaluate concurrence as~\cite{Hill1997,Wootters1998}
	\begin{equation}\label{eq:concpure}
		C_{ij}^{\sf pure} = |\bra{\psi^*} \sigma_{i}^{y}\otimes\sigma_{j}^{y}\ket{\psi}| \; .
	\end{equation}
	
	To see that these two definitions lead to different answers when applied to many--body states 
	with finite multipartite entanglement, we consider a 4--site Heisenberg spin chain 
	with antiferromagnetic interactions ($J > 0$), under periodic boundary conditions
	\begin{equation}
		{\mathcal H} = J ({\bf S}_{1}\cdot {\bf S}_{2} + {\bf S}_{2}\cdot {\bf S}_{3} + {\bf S}_{3}\cdot {\bf S}_{4} + {\bf S}_{4}\cdot {\bf S}_{1}) \; . 
	\end{equation}
	The ground state is of this model is 
	\begin{equation}\label{eq:rvbN=4}
		\ket{\Psi} = \frac{1}{\sqrt{3}} \bigg([12]\otimes[34] + [23]\otimes[41]\bigg) 
	\end{equation} 
	where
	\begin{equation} \label{eq:ground.state}
	[ij] = \frac{1}{\sqrt{2}}(\ket{0_{i}1_{j}} - \ket{1_{i}0_{j}}) \; , 
	\end{equation}
	is a singlet shared between sites $i$ and $j$. 
	 
	 We consider first the pair of sites $1$ and $2$.   Tracing out the spins associated with sites $3$ and $4$, 
	 we find a reduced density matrix describing a mixed state
	 \begin{equation}
	 	\rho_{12} = \begin{pmatrix}
	 						1/12 &0 &0   &0\\
	 						0 &	5/12 &-1/3 &0\\
	 						0 &-1/3  &5/12  &0\\
	 						0 &0 &0 &1/12
 				\end{pmatrix} \; .
	 \end{equation}
	The direct application of the result for a mixed state, Eq.~(\ref{eq:concmixed}), 
	yields a concurrence 
	\begin{equation}
		C_{12}^{\sf mix} = \frac{1}{2} \; , 
	\end{equation}
	reflecting the fact that a singlet, with concurrence $C=1$, 
	and an unentangled state, with concurrence $C=0$, 
	occur with equal probability for this pair of spins in Eq.~(\ref{eq:ground.state}).
	
	If, instead, one were to blindly apply the result for pure state, Eq.~(\ref{eq:concpure}), this would lead to
	an overestimate of the concurrence between the two spins
	\begin{equation}
	 C_{12}^{\sf pure} = \frac{2}{3} \; .
	 \end{equation}
 
	Similarly, for the pair of sites $1$ and $3$, one should find 
	\begin{equation}
	 	C_{13}^{\sf mix} = 0 \;, 
	 \end{equation}
	since the wave function, Eq.~(\ref{eq:ground.state}), does not support any Bell state 
	between this pair of spins.
	In contrast, if we apply the result for a pure state, Eq.~(\ref{eq:concpure}), we find 
	\begin{equation}
		 C_{13}^{\sf pure} = \frac{1}{3}   \;, 
	  \end{equation}
	falsely diagnosing entanglement between this pair of spins.
	 
	 The message, so far as disordered spin chains are concerned, is very clear:
	 random singlet phases involve states with finite multipartite entanglement 
	 (cf. Section~\ref{sec:TLLandRSSMultipartite}).
	 For this reason, it is necessary to work the definition appropriate to a mixed state, Eq.~(\ref{eq:concmixed}),  
	 when calculating the concurrence associated with any given pair of spins.
	 
\end{appendix}

	\bibliography{document.bib}
	
\end{document}